\documentclass[12pt,a4paper]{article}
\usepackage{amsfonts}
\usepackage{amssymb}
\usepackage{latexsym}
\begin{document}

\begin{center}
{\Large \bf Gravity in a stabilized brane world model\\
\vspace{1mm} in five-dimensional Brans-Dicke theory} \\

\vspace{4mm}

A.S.~Mikhailov$^{a}$, Yu.S.~Mikhailov$^{a}$, M.N.~Smolyakov$^{b}$, I.P.~Volobuev$^{b}$\\
\vspace{0.5cm} $^{a}$Department of Physics, Moscow State
University,
\\ 119991, Moscow, Russia\\
$^{b}$Skobeltsyn Institute of Nuclear Physics, Moscow State
University,
\\ 119991, Moscow, Russia\\
\end{center}

\begin{abstract}
Linearized equations of motion for gravitational and scalar fields
are found and solved in a stabilized brane world model in
five-dimensional Brans-Dicke theory. The physical degrees of
freedom are isolated, the mass spectrum of Kaluza-Klein
excitations is found and the coupling constants of these
excitations to matter on the negative tension brane are
calculated.
\end{abstract}

\section{Introduction}
Nowadays models with extra dimensions with the fundamental energy
scale lying in the $TeV$ range are widely discussed in scientific
literature. One of the most known and interesting from the
phenomenological point of view is the Randall-Sundrum model
\cite{Sun}. It describes two branes interacting with gravity in a
five-dimensional space-time and provides an original solution to
the hierarchy problem of gravitational interaction
\cite{Sun,Rubakov,Kubyshin}. Nevertheless, the Randall-Sundrum
model possesses an essential flaw: the distance between the branes
is not fixed by the parameters of the model. This leads to the
existence of the massless scalar field -- the radion -- in the
four-dimensional effective theory on the branes. The coupling
constant of this field to matter on the negative tension brane,
which is assumed to trap the Standard Model fields, appears to be
very large, which contradicts experimental data even at the level
of classical experiments \cite{Rubakov,boos}.

This problem was solved by introducing an extra scalar field
living in the bulk. The most consistent model was proposed in
paper \cite{wolfe}, where exact solutions to equations of motion
for the background metric and scalar field were found. The size of
the extra dimension is defined by the boundary conditions on the
branes. Nevertheless, the background solution for the metric in
this model is more complicated compared to the simple solution in
the Randall-Sundrum model. It turned out that the corresponding
linearized equations of motion in this model can be solved
analytically only for a certain choice of the model parameters,
i.e. when the background solution for the metric can be
approximated by the solution of the unstabilized Randall-Sundrum
model \cite{BVMS}. A question arises: is there a stabilized
solution in the system of two branes admitting the simple
Randall-Sundrum solution for the metric? In paper
\cite{Grzadkowski} it was found that one can obtain such a
solution in the case of non-minimal coupling of stabilizing scalar
field to gravity. But the background solution for the scalar field
in this paper has a rather complicated form.

One of the standard forms of the non-minimal coupling of scalar
field to gravity is the linear interaction with the scalar
curvature used in Brans-Dicke theory. It has been found that it is
possible to get a stabilized model with two branes in
five-dimensional Brans-Dicke theory, admitting the simple
Randall-Sundrum solution for the metric, whereas the solution for
the scalar field also has the form of a simple exponential
function \cite{mmsv}.

In the present paper we study linearized gravity in the stabilized
brane world model proposed in \cite{mmsv}. It turned out that due
to the simplicity of the background solution, the linearized
equations of motion can be solved analytically for any physically
interesting choice of the model parameters, contrary to the case
of stabilized Randall-Sundrum model \cite{wolfe}. We also
calculate the coupling constants of physical degrees of freedom to
matter on the negative tension brane, where the hierarchy problem
of gravitational interaction is solved, and describe the mass
spectrum of Kaluza-Klein modes.

\section{Background solution}
Let us consider gravity in a five-dimensional space-time
$E=M_4\times S^1/Z_2$, interacting with two branes and with the
scalar field $\phi$. Let us denote coordinates in $E$ by
$\{x^M\}=\{x^\mu,y\}$, $ M=0,1,2,3,4$, where $\{x^\mu\},\:
\mu=0,1,2,3$ are four-dimensional coordinates and the coordinate
$y\equiv x^4$, $-L\le y\le L$, corresponds to the extra dimension.
The extra dimension forms the orbifold $S^1/Z_2$, which is a
circle of diameter $2L/\pi$ with the points $y$ and $-y$
identified. Correspondingly, the metric $g_{MN}$ and the scalar
field $\phi$ satisfy the orbifold symmetry conditions
\begin{eqnarray}\label{eq} g_{\mu\nu}(x,-y)=g_{\mu\nu}(x,y),
\quad g_{\mu4}(x,-y)=-g_{\mu4}(x,y), \\ \nonumber
g_{44}(x,-y)=g_{44}(x,y), \quad \phi(x,-y) =\phi(x,y)\ldotp
\end{eqnarray}
The branes are located at the fixed points of the orbifold $y=0$ è
$y=L$.

The action of the model has the form
\begin{eqnarray}\label{s}
S=\int d^4x \int_{-L}^{L}dy \sqrt{-g}\left[\phi
R-\frac{\omega}{\phi}g^{MN}\partial_M\phi
\partial_N\phi-V(\phi)\right]-\\ \nonumber
-\int_{y=0}\sqrt{-\tilde
g}\lambda_1(\phi)d^4x-\int_{y=L}\sqrt{-\tilde
g}\lambda_2(\phi)d^4x.
\end{eqnarray}

Here $V(\phi)$ is the scalar field potential in five-dimensional
space-time, $\lambda_{1,2}(\phi)$ are scalar field potentials on
the branes, $\omega$ is the five-dimensional Brans-Dicke
parameter, $\tilde g_{\mu\nu}$ denotes an induced metric on the
branes. The signature of the metric $g_{MN}$ is chosen to be
$(-,+,+,+,+)$. Subscripts 1 and 2 label the branes.

We consider the following standard form of the background metric
\begin{equation}\label{backgmetric}
ds^2=\gamma_{MN}dx^{M}dx^{N}=e^{2\sigma(y)}\eta_{\mu\nu}dx^\mu
dx^\nu+dy^2
\end{equation}
with $\eta_{\mu\nu}$ being the flat Minkowski metric, which
preserves the Poincar\'e invariance in four-dimensional flat
space-time, and the following form of the background solution for
the scalar field
\begin{equation}\label{backgscalar}
\phi(x,y)=\phi(y).
\end{equation}
Functions $\sigma(y), \phi(y)$ are defined by the equations of
motion \cite{mmsv}
\begin{eqnarray}\label{backgreq1}
\frac{\omega}{\phi}(\phi''+4\sigma'\phi')
-4\sigma''-10(\sigma')^2&-&\\ \nonumber
-\frac{\omega}{2\phi^2}(\phi')^2-\frac{1}{2}\frac{dV}{d\phi}-
\frac{1}{2}\frac{d\lambda_1 }{d\phi}\delta(y)
-\frac{1}{2}\frac{d\lambda_2 }{d\phi}\delta(y-L)&=&0, \\
\label{backgreq2}
 6(\sigma')^2\phi-\frac{1}{2}\left(\frac{\omega}{\phi} (\phi')^2-V\right)+4\sigma'\phi'&=&0, \label{yd}
 \\ \label{backgreq3}
3\sigma''\phi+\frac{\omega}{\phi}(\phi')^2+\phi''-\sigma'\phi'+
 \frac{1}{2}\lambda_1\delta(y)+\frac{1}{2}\lambda_2\delta(y-L)&=&0.
\end{eqnarray}
Here and below $'=\partial_{4}={\partial}/{\partial{y}}$.

Let us consider the scalar field potentials to be
\begin{eqnarray}\label{super}
& V(\phi)=\Lambda \phi,\qquad \lambda_{1,2}=\pm\lambda \phi.
\end{eqnarray}
In this case functions $\sigma(y)$, $\phi(y)$, which are solution
to equations (\ref{backgreq1})-(\ref{backgreq3}), take the form
\begin{eqnarray}\label{fon}
\sigma=-k|y|+C, \qquad \phi=C_{1}e^{-u|y|},
\end{eqnarray}
where
\begin{eqnarray}\label{consts-bck}
\nonumber&u=\sqrt{\frac{-\Lambda}{(3\omega+4)(4\omega+5)}}, \\
 &k=(\omega+1)u,\\
\nonumber&
\lambda=4\sqrt{-\Lambda}\sqrt{\frac{3\omega+4}{4\omega+5}}.
\end{eqnarray}
Background solution for the metric has the same form as that in
the unstabilized Randall-Sundrum model \cite{Sun}. We consider the
case when the matter is located on the brane at $y=L$, like in the
Randall-Sundrum model. To this end we take $C=kL$, which makes the
four-dimensional coordinates on this brane Galilean (see
\cite{Rubakov,Kubyshin,boos}).

To fix the size of the extra dimension, let us add the following
terms to the brane potentials:
\begin{eqnarray}
\Delta\lambda_{1,2}=\frac{\beta_{1,2}^{2}}{2}(\phi-v_{1,2})^2.
\end{eqnarray}
The equations of motion appear to be satisfied provided
\begin{eqnarray}
\phi|_{y=0}=C_{1}=v_1,\qquad \phi|_{y=L}=v_2.
\end{eqnarray}
Thus, the distance between the branes is defined by the boundary
conditions for the field $\phi$ and can be expressed through the
parameters of the potentials as
\begin{equation}
L= \frac{1}{u} \ln\left(\frac{v_1}{v_2}\right).
\end{equation}
We suppose that $v_{1}\simeq v_{2}$ and $uL<1$. We also suppose
that $kL\approx 35$, which can be achieved if $\omega>35$. Note
that the mechanism of stabilization is based on the dependence of
the scalar field background solution on the coordinate of the
extra dimension. Parameters $v_{1,2}$ of the potentials, made
dimensionless by the fundamental five-dimensional energy scale
$\Lambda$, should be positive values of the order of $O(1)$, i.e.
there should not be any hierarchical difference.

As it was noted above, the five-dimensional scalar field usually
minimally couples to gravity in stabilized brane world models
\cite{wolfe}. In the case of non-minimal coupling, there exists a
conformal transformation transforming the action from the Jordan
frame to the Einstein frame, in which scalar field minimally
couples to gravity. For the action of form (\ref{s}) these
transformations are presented in \cite{mmsv} (with the coordinate
transformations bringing the background metric to the standard
form). The difference between the initial and the transformed
actions is characterized by the interaction of matter on the
branes with gravity, i.e. by the metric which we are supposed to
perceive. It appears that if we live in the world, in which
five-dimensional scalar field is non-minimally coupled to
five-dimensional curvature, it is more convenient to examine
linearized gravity using the initial, untransformed action. It
allows one to simplify the derivation of the Kaluza-Klein mode
mass spectrum and the coupling constants of these modes to matter
on the branes. But the use of the conformal transformations allows
one to simplify the choice of gauge conditions, which are
necessary for isolating the physical degrees of freedom of the
theory. We will discuss this point in the next section.

\section {Linearized equations of motion and the choice of gauge conditions}
To study linearized gravity one should derive the second variation
Lagrangian of the model. To this end let us parameterize the
metric and the scalar field as
\begin{equation}\label{razlozh}
g_{MN}(x,y)=\gamma_{MN}(y)+h_{MN}(x,y),\quad
\phi(x,y)=\phi(y)+f(x,y),
\end{equation}
where $\phi(y)$ is the background solution of the scalar field,
substitute it into action (\ref{s}) and retain the terms of the
second order in fluctuations (below we will use the short notation
$\phi$ for the background solution $\phi(y)$). The corresponding
second variation Lagrangian appears to be extremely large and we
do not present it here in the explicit form. We only present the
linearized equations of motion for fluctuations of metric and
stabilizing scalar field, which follow from this Lagrangian:
\begin{enumerate}
\item $\mu\nu$-component
\begin{eqnarray}\label{eqmunu}
&-\frac{1}{2}\left[\phi\left(\partial_\sigma{\partial^\sigma{h_{\mu\nu}}}-\partial_\mu
{\partial^\sigma{h_{\sigma\nu}}}-\partial_\nu{\partial^\sigma{h_{\sigma\mu}}}
+\partial_4{\partial_4{h_{\mu\nu}}}\right)+\phi\partial_\mu{\partial_
\nu{\tilde{h}}}+\right.
\\ \nonumber
&+\phi\partial_\mu{\partial_\nu{h_{44}}}-\phi\partial_4{
(\partial_\mu{h_{4\nu}}+\partial_\nu{h_{4\mu}})}
-2\sigma'\phi(\partial_\mu{h_{4\nu}}+
\partial_\nu{h_{4\mu}})+\\ \nonumber
& +\phi\gamma_{\mu\nu}\biggl( -\partial_4 {\partial_4 {\tilde{h}}}
-\partial_{\sigma}{\partial^{\sigma}
{h_{44}}}-\partial_\sigma{\partial^\sigma
{\tilde{h}}}-4\sigma'\partial_4{\tilde{h}}+3\sigma'\partial_4{h_{44}}
+\\ \nonumber
&+\partial_{\sigma}{\partial_{\tau}{h^{\sigma\tau}}}+
2\partial^{\sigma}{\partial_4{h_{\sigma
4}}}+4\sigma'\partial^\sigma{h_{4 \sigma}}
\biggr)-2h_{\mu\nu}(2(\sigma')^2\phi+\sigma''\phi+\sigma'\phi')+\\
\nonumber &+3h_{44}\gamma_{\mu\nu}(4(\sigma')^2+\sigma'')-
\phi'(\partial_{\mu}h_{\nu4}+\partial_{\nu}h_{\mu4}-\partial_{4}h_{
\mu\nu})+2\partial_\mu\partial_\nu f+\\ \nonumber
&+\gamma_{\mu\nu}\phi'\biggl(\partial_{4}h_{44}+2\partial^{\sigma}h_{\sigma4}-\partial_{4}\tilde
h+7\sigma'h_{44}-2\frac{\omega}{\phi}f'-8\frac{\omega}{\phi}f\sigma'\biggr)
+\\ \nonumber
&+\gamma_{\mu\nu}\biggl(h_{44}\phi''-2f''-2\partial_\sigma\partial^\sigma
f+2f\sigma''+\\ \nonumber
&\left.+8f(\sigma')^2+2\frac{\omega}{\phi^2}f(\phi')^2-2\frac{\omega}{\phi}f\phi''-6\sigma'f'
\biggr)\right]=0,
\end{eqnarray}

\item $\mu 4$-component
\begin{eqnarray}\label{hm4}
&\frac{1}{2}\left[\phi\partial_4(\partial_\mu\tilde{h}-\partial^{\nu}
h_{\mu\nu})+\phi\partial^\nu(\partial_\nu h_{\mu 4 }-\partial_\mu
h_{\nu 4})-3\sigma'\phi\partial_{\mu} h_{44}+\right.\\ \nonumber
&\left.+ 2\partial_\mu\partial_4f-2\sigma'\partial_\mu
f-\phi'\partial_\mu h_{44}+2\frac{\omega}{\phi}\partial_\mu
f\phi'\right]=0,&
\end{eqnarray}
\item $44$-component
\begin{eqnarray}\label{eq44}
&-\frac{1}{2}\left[\phi\partial^\mu(\partial^\nu
h_{\mu\nu}-\partial_{\mu} \tilde{h})+6\phi \sigma'\partial^{\mu}
h_{\mu4} -3\phi\sigma'\partial_4{\tilde{h}}-12f(\sigma')^2+\right. \\
\nonumber &+12\phi h_{44}(\sigma')^2 +2\phi'(\partial^\mu
h_{\mu4}+4\sigma'h_{44}-\frac{1}{2}\partial_4\tilde
h)-2\partial_\mu\partial^\mu
f-\\
\nonumber
&\left.-8\sigma'f'-\frac{\omega}{\phi^2}(\phi')^2f+2\frac{\omega}{\phi}\phi'f'
-\frac{\omega}{\phi}(\phi')^2h_{44}-\frac{dV}{d\phi}f\right] =0,&
\end{eqnarray}
\item equation for the field $f$
\begin{eqnarray}\label{eqf}
&\partial^\mu(\partial^\nu h_{\mu\nu}-\partial_\mu
h_{44}-\partial_\mu \tilde h+2\partial_4
h_{\mu4})-\partial_4\partial_4 \tilde h+
\\ \nonumber
&+\partial^\mu
h_{\mu4}(10\sigma'-2\frac{\omega}{\phi}\phi')+\partial_4h_{44}(4\sigma'-\frac{\omega}{\phi}\phi')+\partial_4\tilde
h(-5\sigma'+\frac{\omega}{\phi}\phi')-
\\ \nonumber
&-h_{44}\biggl(\frac{d V }{d \phi}+\frac{1}{2}\frac{d \lambda_1
}{d\phi}\delta(y)+\frac{1}{2}\frac{d \lambda_2 }{d\phi}\delta(y-L)
\biggr)+\\
\nonumber &+f\biggl(
2\frac{\omega}{\phi^3}(\phi')^2-8\frac{\omega}{\phi^2}\sigma'\phi'
-2\frac{\omega}{\phi^2}\phi''-\frac{d^2 V }{d \phi^2}-\frac{d^2
\lambda_1 }{d\phi^2}\delta(y)-\\ \nonumber &-\frac{d^2 \lambda_2
}{d\phi^2}\delta(y-L) \biggr)+f'\left(
8\frac{\omega}{\phi}\sigma'-2\frac{\omega}{\phi^2}\phi'\right)
+2\frac{\omega}{\phi}f''+2\frac{\omega}{\phi}\partial^\mu\partial_\mu
f=0,&
\end{eqnarray}
\end{enumerate}
where $\; h=\gamma_{MN}h^{MN}, \;
\tilde{h}=\gamma_{\mu\nu}h^{\mu\nu}$. Note that these equations
can also be obtained by linearizing the Einstein equations and the
equation for the scalar field, which follow from action (\ref{s}),
in the background defined by solution of form (\ref{backgmetric}),
(\ref{backgscalar}).

Let us now discuss the gauge invariance of the linearized theory.
It is not difficult to check that the quadratic action is
invariant under gauge transformations of the form
\begin{eqnarray*}
h_{MN}^{(\prime)}&=&h_{MN}-(\nabla_M\xi_N+ \nabla_N\xi_M),\\
f^{(\prime)}&=& f-\phi'\xi_4,
\end{eqnarray*}
where $\nabla_M$ denotes the covariant derivative with respect to
the background metric $\gamma_{MN}$, if functions $\xi^M$ satisfy
the orbifold symmetry conditions
\begin{equation}\label{sim}
\xi^\mu(x,-y)=\xi^\mu(x,y),\qquad\xi^4 (x,-y)=-\xi^4(x,y)
\end{equation}
(here $(\prime)$ denotes the transformed field). The existence of
these gauge transformations is a consequence of the invariance of
action (\ref{s}) under the general coordinate transformations.
Analogous gauge transformations were discussed in the case of
Randall-Sundrum model without the stabilizing scalar field and in
the case of stabilized Randall-Sundrum model
\cite{boos,BVMS,Aref,11}. One can use them to isolate the physical
degrees of freedom of the fields $h_{MN}$ and $f$. Though this
problem can be simplified. Indeed, for the case of the minimal
coupling of the stabilizing scalar field there were found gauge
conditions, which allows one to isolate the physical degrees of
freedom in the general case \cite{BVMS}. Using the conformal
transformations, supplemented by the coordinate transformations
\cite{mmsv}, one can obtain from the gauge conditions,
corresponding to the minimal coupling of the scalar field, the
gauge conditions, corresponding to action (\ref{s}) and background
metric (\ref{backgmetric}):
\begin{eqnarray}\label{kal}
&\partial_4
\left[\left(h_{44}+\frac{2}{3}\frac{f}{\phi}\right)e^{2\sigma}\phi^{2/3}\right]=
\frac{4}{3}\left(\omega+\frac{4}{3}\right)e^{2\sigma}\frac{\phi'}{\phi^{4/3}}f,\\
\nonumber & h_{\mu4}=0.&
\end{eqnarray}
Analogously we can find the substitution, which allows us to
diagonalize equations of motion (\ref{eqmunu})-(\ref{eqf}), as
well as the second variation Lagrangian:
\begin{eqnarray}\label{podst}
h_{\mu\nu}=b_{\mu\nu}-\frac{1}{2}\gamma_{\mu\nu}h_{44}-\gamma_{\mu\nu}\frac{f}{\phi},
\end{eqnarray}
where $b_{\mu\nu}$ is a traceless-transverse field.

In gauge (\ref{kal}) and with substitution (\ref{podst}) equation
(\ref{hm4}) fulfills automatically. Equation (\ref{eqmunu}) takes
the form
\begin{equation}\label{ub}
\frac{1}{2}\left(\partial_\sigma{\partial^\sigma{b_{\mu\nu}}}
+\frac{\partial^2}{\partial y^2}
{b_{\mu\nu}}\right)-b_{\mu\nu}\left
(2(\sigma')^2+\sigma''+\sigma'\frac{\phi'}{\phi}\right)+\frac{\phi'}{2\phi}b_{\mu\nu}'=0.
\end{equation}

Equation for 44-component (\ref{eq44}) simplifies considerably, if
one rewrites it in the interval $(0,L)$ using a new function $g =
e^{2\sigma(y)}\phi^{2/3}(h_{44}(x,y)+\frac{2}{3}\frac{f}{\phi})$
and taking into account the relation between the potential $V$ and
functions $\sigma$, $\phi$ (\ref{yd}):
\begin{equation}\label{ug0}
g'' +g'\left(\frac{5}{3}\frac{\phi'}{\phi}
-2\sigma'-2\frac{\phi''}{\phi'}\right)-\frac{2}{3}\frac{(\phi')^2}{(\phi)^2}\left(\omega+\frac{4}{3}\right)
g+\partial_\mu \partial^\mu g =0.
\end{equation}
In terms of function $g$, substitution (\ref{podst}) and gauge
conditions (\ref{kal}) take the form:
\begin{eqnarray}\label{subst}
h_{\mu\nu} &=& b_{\mu\nu} -
\frac{1}{2}\gamma_{\mu\nu}e^{-2\sigma}\phi^{-2/3}g-\frac{2}{3}\gamma_{\mu\nu}\frac{f}{\phi},
\\ \label{subst11}
 h_{44} &=& e^{-2\sigma}\phi^{-2/3} g-\frac{2}{3}\frac{f}{\phi},
\\ \label{subst22}
g'&=&\frac{4}{3}\left(\omega+\frac{4}{3}\right)e^{2\sigma}\frac{\phi'}{\phi^{4/3}}f,
\\ \label{subst33}
 h_{\mu 4} &=& 0, \quad  \tilde b = \gamma_{\mu\nu} b^{\mu\nu} =0, \quad
\partial^\nu{b_{\mu\nu}}=0.
\end{eqnarray}
Note, that the field $g$ is a superposition of $h_{44}$ component
of the metric fluctuations and the fluctuation $f$ of the
stabilizing scalar field.

Substitution of (\ref{subst})-(\ref{subst33}) into equation
(\ref{eqf}) results in the equation, which can be obtained by
differentiating (\ref{ug0}) by $y$, and in the boundary conditions
on the branes:
\begin{eqnarray}\label{bc}
&\left(
\frac{\phi''}{\phi'}-\frac{1}{4}\frac{\phi}{(\omega+4/3)}\frac{d^2
\lambda_1}{d\phi^2}
+\frac{1}{(3\omega+4)}(\omega\frac{\phi'}{\phi}-4\sigma')
\right)g' -\partial_\mu
\partial^\mu g|_{y=+0}=0,
&\\ \nonumber &\left(
\frac{\phi''}{\phi'}+\frac{1}{4}\frac{\phi}{(\omega+4/3)}\frac{d^2
\lambda_2}{d\phi^2}
+\frac{1}{(3\omega+4)}(\omega\frac{\phi'}{\phi}-4\sigma')
\right)g' -\partial_\mu
\partial^\mu g|_{y=L-0}=0.&
\end{eqnarray}
For the case of background solution (\ref{fon}),
(\ref{consts-bck}) the boundary conditions simplify considerably
and take the form
\begin{eqnarray}\label{bc1}
&\frac{1}{4}\frac{\phi}{(\omega+4/3)}\beta_{1}^{2}g' +\partial_\mu
\partial^\mu g|_{y=+0}=0,
&\\ \nonumber &\frac{1}{4}\frac{\phi}{(\omega+4/3)}\beta_{2}^{2}g'
-\partial_\mu\partial^\mu g|_{y=L-0}=0.&
\end{eqnarray}
It should be noted that such a simplification of the boundary
conditions takes place for a class of background solutions in
five-dimensional Brans-Dicke theory, namely, if the equations of
motion for the background configuration of the fields can be
reduced to first order differential equations \cite{mmsv}.

\section{Mass spectrum of Kaluza-Klein modes and\\ four-dimensional effective Lagrangian}
First let us consider the tensor modes of the field $b_{\mu\nu}$,
which satisfies equation (\ref{ub}). To find the mass spectrum and
wave functions in the extra dimension, let us represent the field
$b_{\mu\nu}$ as
\begin{equation}\label{decomp-t}
b_{\mu\nu}(x,y)=\sum_{n=0}^{\infty}b_{\mu\nu}^{n}(x)\psi_{n}(y),\quad
\Box b_{\mu\nu}^{n}(x)=m_{n}^{2}b_{\mu\nu}^{n}(x),
\end{equation}
where $\Box=\eta^{\mu\nu}\partial_{\mu}\partial_{\nu}$.
Substituting this into (\ref{ub}), we obtain
\begin{equation}\label{ub1}
e^{-2\sigma}m_{n}^{2}\psi_{n} +\psi_{n}''-\left
(4(\sigma')^2+2\sigma''+2\sigma'\frac{\phi'}{\phi}\right)\psi_{n}+\frac{\phi'}{\phi}\psi_{n}'=0.
\end{equation}
The boundary conditions following from this equation can be
obtained by integrating (\ref{ub1}) in an infinitely small
vicinity of the points $y=0$, $y=L$ and have the form
\begin{eqnarray}
\psi_{n}'-2\sigma'\psi_{n}|_{y=+0}=0,\\
\psi_{n}'-2\sigma'\psi_{n}|_{y=L-0}=0.
\end{eqnarray}
It follows from the general theory \cite{BKM} that all eigenvalues
of the problem under consideration are real and positive. Thus,
the tensor sector does not contain tachyons. The eigenfunctions
corresponding to different eigenvalues are orthogonal with the
weight function defined by equation (\ref{ub1}). The
eigenfunctions can be normalized as follows
\begin{equation}\label{norm-ub}
\int_{-L}^{L}\phi e^{-2\sigma}\psi_{n}\psi_{k}dy=\delta_{nk}.
\end{equation}

The solution for the zero mode is
\begin{equation}\label{zeromode}
\psi_0(y)=
\sqrt{\frac{k+u/2}{v_{1}}}\frac{e^{-kL}}{\sqrt{1-e^{-2kL-uL}}}\,e^{2kL-2k|y|},
\end{equation}
and if $u\ll k$ and $kL\approx 35$
\begin{equation}\label{zeromode-brane}
\psi_0(L)\approx \sqrt{\frac{k}{v_{1}}}\,e^{-kL}.
\end{equation}

For the case of massive modes equation (\ref{ub1}) in the interval
$(0,L)$ can be solved in a standard way (see, for example,
\cite{boos,BVMS}) by passing to variable
$z=\frac{m_{n}}{k}e^{ky-kL}$, which leads to
\begin{equation}
\psi_n(z)=z^{a}\left(AJ_{\alpha}(z)+BJ_{-\alpha}(z)\right),
\end{equation}
where $J_{\alpha}(z)$ is the Bessel function, $a=\frac{u}{2k}$,
$\alpha=2+\frac{u}{2k}$. In the next section we will show that one
should take $kL \approx 35$ to obtain the (weak) Newtonian gravity
on the brane at $L$ retaining a strong five-dimensional gravity.
In this case one can use the approximation ${z|_{y=0}
=\frac{m_n}{k} e^{-kL}}\approx 0$ with a good accuracy, which
allows one to drop the singular term $J_{-\alpha}(z)$ in
$\psi_n(z)$, because $B/A \sim e^{-2\alpha kL}$ and the
corresponding corrections are negligible. The boundary condition
at $L$ gives us the mass spectrum of tensor Kaluza-Klein modes,
which is defined by
\begin{equation}\label{spectrum-t}
J_{\alpha-1}\left(\frac{m_{n}}{k}\right)=0.
\end{equation}
Note that in the limit $u=0$ we reproduce the equation for the
mass spectrum of tensor modes in the unstabilized Randall-Sundrum
model $J_{1}\left(\frac{m_{n}}{k}\right)=0$ \cite{boos}.
Normalization constant $A$ is defined by formula (\ref{norm-ub}),
and the normalized wave functions of massive tensor modes look as
follows:
\begin{equation}\label{massive-norm}
\psi_n(y)=\sqrt{\frac{k}{v_{1}}}e^{\frac{u|y|}{2}}\frac{J_{\alpha}\left(\frac{m_{n}}{k}e^{k|y|-kL}\right)}{J_{\alpha}\left(\frac{m_{n}}{k}\right)}.
\end{equation}
At the point $y=L$ we get a simple formula
\begin{equation}\label{massive-norm-L}
\psi_n(L)=\sqrt{\frac{k}{v_{2}}}.
\end{equation}

Now let us turn to the scalar sector. To find the mass spectrum of
the scalar modes defined by equation (\ref{ub}) we represent $g$
as
$$
g(x,y) = e^{ipx} g_n(y), \quad p^2 = -\mu_n^2.
$$
As a result equation (\ref{ug0}) and boundary conditions
(\ref{bc1}) for $ g_n(y)$ take the form
\begin{equation}\label{ug-modes}
g_{n}'' +g_{n}'\left(\frac{5}{3}\frac{\phi'}{\phi}
-2\sigma'-2\frac{\phi''}{\phi'}\right)-\frac{2}{3}\frac{(\phi')^2}{(\phi)^2}\left(\omega+\frac{4}{3}\right)
g_{n}+e^{-2\sigma}\mu_n^2 g_{n}=0,
\end{equation}
\begin{eqnarray}\label{bc-modes}
&\frac{1}{4}\frac{\phi}{(\omega+4/3)}\beta_{1}^{2}g_{n}'
+e^{-2\sigma}\mu_n^2 g_{n}|_{y=+0}=0, &\\ \label{bc-modes1}
&\frac{1}{4}\frac{\phi}{(\omega+4/3)}\beta_{2}^{2}g_{n}'
-e^{-2\sigma}\mu_n^2 g_{n}|_{y=L-0}=0.&
\end{eqnarray}
Equation (\ref{ug-modes}) is written in the interval $(0,L)$, but
it can be combined with the boundary conditions (\ref{bc-modes}),
(\ref{bc-modes1}), which results in a single equation on the
circle $S^1$:
\begin{eqnarray}\label{ug-S}
\left(g_{n}'\frac{\phi^{5/3}e^{-2\sigma}}{\phi'^{2}}\right)'
-\frac{2}{9}\left(3\omega+4\right)\frac{e^{-2\sigma}}{\phi^{1/3}}
g_{n}+\\
\nonumber
+\frac{\phi^{5/3}e^{-4\sigma}}{\phi'^{2}}\left(\frac{8(3\omega+4)}{3\beta_{1}^{2}\phi}\delta(y)+
\frac{8(3\omega+4)}{3\beta_{2}^{2}\phi}\delta(y-L)+1\right)\mu_n^2
g_{n} =0.
\end{eqnarray}
With the help of this equation one can show that
\begin{equation}
\int_{-L}^{L}dy\left(\frac{\phi^{5/3}e^{-2\sigma}}{\phi'^{2}}g'_{n}g'_{k}+\frac{2}{9}\left(3\omega+4\right)\frac{e^{-2\sigma}}{\phi^{1/3}}
g_{n}g_{k}\right)=0
\end{equation}
for $n\ne k$ (we suppose that $\mu_n\ne \mu_k$),
\begin{eqnarray}\label{nozero}
\int_{-L}^{L}dy\left(\frac{\phi^{5/3}e^{-2\sigma}}{\phi'^{2}}{g'}_{n}^{2}+\frac{2}{9}\left(3\omega+4\right)\frac{e^{-2\sigma}}{\phi^{1/3}}
g_{n}^{2}\right)=\\ \nonumber
=\mu_{n}^{2}\int_{-L}^{L}dy\frac{\phi^{5/3}e^{-4\sigma}}{\phi'^{2}}\left(\frac{8(3\omega+4)}{3\beta_{1}^{2}\phi}\delta(y)+
\frac{8(3\omega+4)}{3\beta_{2}^{2}\phi}\delta(y-L)+1\right)g_{n}^{2}.
\end{eqnarray}
From (\ref{nozero}) it follows that the scalar sector has no zero
mode with $\mu_{0}=0$, because if $\mu_{0}=0$, then
$g_{0}(y)\equiv 0$. If $\beta_{1,2}^{2}>0$, then $\mu_{n}^{2}>0$,
i.e. the scalar sector does not contain tachyons. Using the
formulas presented above, we will normalize the wave functions as
\begin{eqnarray}\label{norm}
\mu_{n}^{2}\int_{-L}^{L}dy\frac{\phi^{5/3}e^{-4\sigma}}{\phi'^{2}}\left(\frac{8(3\omega+4)}{3\beta_{1}^{2}\phi}\delta(y)+
\frac{8(3\omega+4)}{3\beta_{2}^{2}\phi}\delta(y-L)+1\right)g_{n}g_{k}=\\
\nonumber=\frac{8(3\omega+4)}{27}\delta_{nk}.
\end{eqnarray}
Five-dimensional scalar field $g(x,y)$ can be represented as a
series
\begin{equation}\label{decomps}
g(x,y) =\sum_{n=1}^\infty \varphi_n(x)g_n(y),
\end{equation}
where four-dimensional scalar fields $\varphi_n(x)$ have the
masses $\mu_n$.

With (\ref{fon}) and (\ref{consts-bck}) equation (\ref{ug-modes})
in the interval $(0,L)$ takes the form
\begin{equation}\label{ug-modes-fon}
g_{n}''
+g_{n}'\left(2k+\frac{u}{3}\right)-\frac{2}{3}u^{2}\left(\omega+\frac{4}{3}\right)
g_{n}+e^{2ky-2kL}\mu_n^2 g_{n}=0.
\end{equation}
It can be solved by passing to the variable
$z=\frac{\mu_{n}}{k}e^{ky-kL}$. After some calculations (which are
absolutely equivalent to those carried out in \cite{BVMS} for the
scalar sector of stabilized Randall-Sundrum model), we obtain
\begin{equation}\label{scalsol}
g_{n}(z)=z^{q}\left(A_{n}J_{\gamma}(z)+B_{n}J_{-\gamma}(z)\right),
\end{equation}
with $$q=-1-\frac{u}{6k}=-\frac{6\omega+7}{6\omega+6},$$
$$\gamma=\sqrt{\frac{2u^{2}}{3k^{2}}\left(\omega+\frac{4}{3}\right)+q^{2}}=\frac{2\omega+3}{2\omega+2}.$$
Substituting (\ref{scalsol}) into the boundary condition at zero,
we can drop the singular term $\sim J_{-\gamma}(z)$ (as for the
tensor sector), i.e. $B_{n}=0$. The boundary condition at $L$
defines the mass spectrum and takes the form
\begin{equation}\label{spect-scal}
J_{\gamma}\left(\frac{\mu_{n}}{k}\right)\left[1+\frac{v_{2}\beta_{2}^{2}k}{\mu_{n}^{2}(2\omega+2)}\right]=
J_{\gamma-1}\left(\frac{\mu_{n}}{k}\right)\frac{3v_{2}\beta_{2}^{2}}{4\mu_{n}(3\omega+4)}.
\end{equation}

The wave functions $g_{n}(y)$ take the form
\begin{equation}\label{func-scal-n}
g_{n}(y)=A_{n}\left(\frac{\mu_{n}}{k}e^{k|y|-kL}\right)^{q}J_{\gamma}\left(\frac{\mu_{n}}{k}e^{k|y|-kL}\right),
\end{equation}
\begin{eqnarray}\label{coeff-scal-n}
&&A_{n}=\frac{u}{3\mu_{n}J_{\gamma}\left(\frac{\mu_{n}}{k}\right)}\left(\frac{\mu_{n}}{k}\right)^{-q}\times\\
\nonumber &&\times
\left[\frac{3}{8v_{2}^{1/3}k(3\omega+4)}\left(1-\frac{\gamma^{2}k^{2}}{\mu_{n}^{2}}+\left(
\frac{4(3\omega+4)\mu_{n}}{3v_{2}\beta_{2}^{2}}-\frac{qk}{\mu_{n}}\right)^{2}\right)+\frac{1}{\beta_{2}^{2}v_{2}^{4/3}}\right]^{-\frac{1}{2}},
\end{eqnarray}
where the normalization coefficient $A_{n}$ is derived from
(\ref{norm}).

To get an effective four-dimensional Lagrangian of the theory
(which also allows one to check the correctness of normalization
conditions (\ref{norm-ub}) and (\ref{norm}) by checking the
coefficients in front of the four-dimensional kinetic terms of the
fields), we need the second variation Lagrangian of the model. The
necessary part of the second variation Lagrangian is
\begin{eqnarray}\label{l}
L_g/\sqrt{-\gamma}=\frac{1}{2}\Bigl(-h^{\mu\nu}\times\left[\textnormal{eq.
(\ref{eqmunu})}\right]_{\mu\nu}-h^{44}\times\left[\textnormal{eq.
(\ref{eq44})}\right]+f\times\left[\textnormal{eq.
(\ref{eqf})}\right]\Bigr),
\end{eqnarray}
where we have taken into account $h_{\mu 4}\equiv 0$, and
$\left[\textnormal{eq. (\ref{eqmunu})}\right]_{\mu\nu}$,
$\left[\textnormal{eq. (\ref{eq44})}\right]$,
$\left[\textnormal{eq. (\ref{eqf})}\right]$ are the left hand
sides of equations (\ref{eqmunu}), (\ref{eq44}), (\ref{eqf})
respectively. This result is not surprising: in fact, formula
(\ref{l}) follows from the definition of the second variation
Lagrangian. Thus, in principle, one can obtain the second
variation Lagrangian using only linearized equations of motion for
the fields.

Substituting (\ref{decomp-t}) and (\ref{decomps}) into second
variation Lagrangian (\ref{l}), taking into account
(\ref{subst})-(\ref{subst33}), (\ref{norm-ub}), (\ref{norm}) and
integrating over the extra dimension, we get
\begin{eqnarray}
S_{eff}=-\frac{1}{4}\sum_{k=0}^\infty\int d^{4}x \left(
\partial^\sigma b^{k,\mu \nu}\partial_\sigma b_{\mu \nu}^k+m_k^2
b^{k,\mu \nu}b_{\mu \nu}^k\right)-\frac{1}{2}\sum_{k=1}^\infty\int
d^{4}x \left(\partial_\nu \varphi_k\partial^\nu \varphi_k+\mu_k^2
\varphi_k \varphi_k \right).
\end{eqnarray}
Thus, we have obtained the effective Lagrangian of the theory,
which is the sum of the standard four-dimensional Lagrangians for
the scalar and tensor fields. We note that the kinetic terms in
the effective Lagrangian have the proper sign, i.e. there are no
phantom fields in the four-dimensional effective theory.

\section{Interaction with matter}
Interaction of four-dimensional fields $b_{\mu\nu}^n(x)$ and
$\varphi_n(x)$ with the Standard Model fields on the branes is
described by the interaction of the fluctuations of
five-dimensional gravitational field $h_{\mu\nu}$ with matted on
the branes, which has the standard form
\begin{eqnarray}\label{vz}
\frac{1}{2}\int_{B_1}h_{\mu\nu}(x,0)T_{(1)}^{\mu\nu}
\sqrt{-det\gamma_{\rho\sigma}(0)}d^{4}x+\frac{1}{2}
\int_{B_2}h_{\mu\nu}(x,L)T_{(2)}^{\mu\nu}\sqrt{-det\gamma_{\rho\sigma}(L)}d^{4}x,
\end{eqnarray}
$T_{(1)}^{\mu\nu}$ and $T_{(2)}^{\mu\nu}$ being energy-momentum
tensors of matter on brane~1 and brane~2 respectively.

We restrict ourselves to matter on the brane at $y=L$, which is
supposed to be "our"\, brane. Substituting expansions
(\ref{decomp-t}) and (\ref{decomps}) into (\ref{vz}), taking into
account (\ref{subst}), (\ref{subst11}), (\ref{subst22}) and
(\ref{bc-modes1}), we get the formula describing interaction of
tensor and scalar modes with matter on the brane at $y=L$ in
Galilean coordinates on that brane
\begin{eqnarray}\label{allint}
\frac{1}{2}\int_{B_2}\left(\psi_0(L)b_{\mu\nu}^0(x)T^{\mu\nu}
+\sum_{n=1}^\infty\psi_n(L)b_{\mu\nu}^n(x)T^{\mu\nu} -\right.\\
\nonumber
\left.-\frac{1}{2}v_{2}^{-2/3}\sum_{n=1}^\infty\left(1-\frac{4\mu_{n}^{2}}{uv_{2}\beta_{2}^{2}}\right)
g_n(L) \varphi_n(x) T_\mu^\mu\right)d^{4}x.
\end{eqnarray}

The coefficient in front of the zero tensor mode $b_{\mu\nu}^0(x)$
defines the four-dimensional Planck mass on the brane, and with
the use of (\ref{zeromode-brane}) we obtain
\begin{equation}
M_{Pl}=\psi_0^{-1}(L)\approx \sqrt{\frac{v_{1}}{k}}\,e^{kL}.
\end{equation}
If $\sqrt{\frac{k}{v_{1}}}\sim 1\,\textit{TeV}^{-1}$, whereas
$kL\approx 35$, then $M_{Pl}\sim 10^{19}\textit{GeV}$. Thus, the
hierarchy problem of gravitational interaction on the brane at
$y=L$ is solved analogously to that in the Randall-Sundrum model.
The coupling constants to matter on the brane have the form
\begin{equation}
\frac{\psi_n(L)}{2}\approx \sqrt{\frac{k}{4v_{2}}}\sim
1\,\textit{TeV}^{-1},
\end{equation}
where we have used formula (\ref{massive-norm-L}). When
$\omega>35$, one gets $\alpha\approx 2$ and the mass spectrum of
the tensor modes is approximately the same as that in the
unstabilized Randall-Sundrum model \cite{boos}. In this case
$m_{1}\approx 3.8k$ and can be of the order of
$3-4\,\textit{TeV}$.

Now let us turn to the scalar sector. We would like to mention an
interesting feature of the coupling constants: in principle it is
possible that for appropriate values of the parameters
$$\mu_{j}=\sqrt{\frac{uv_{2}\beta_{2}^{2}}{4}}$$
for some $j$. In this case $j$-th mode of the scalar field does
not interact with matter on the brane.

Now let us estimate the mass and the coupling constant of the
lightest scalar mode, -- the radion. To simplify the analysis we
use the "stiff brane potential"\, limit --
$\beta_{2}^{2}\to\infty$. Let us suppose that $\mu_{1}<k$.
Expanding the Bessel functions in (\ref{spect-scal}) into a
series, retaining the terms up to the second order in
$\frac{\mu_{1}}{k}$ and solving the resulting quadratic equation,
we get
\begin{equation}\label{radmass}
\mu_{1}\approx\frac{2k}{\sqrt{\omega}}.
\end{equation}
Since $\omega>35$, then $\mu_{1}<k$, which confirms the validity
of the expansions of the Bessel functions.

Note, that if $\beta_{2}^{2}\to\infty$, then ${g'}_{n}(L)=0$,
which follows from the boundary condition (\ref{bc-modes1}).
Thereby, only the first one of the two terms in (\ref{allint}),
describing the interaction of the scalar modes with matter on the
brane, remains, and the coupling constants look like
\begin{equation}
\epsilon_n=-\frac{1}{4}v_{2}^{-2/3}g_n(L).
\end{equation}
Taking into account (\ref{radmass}) and $\omega>35$, for the
lightest mode -- the radion -- this constant is simply
\begin{equation}
\epsilon_1\approx-\frac{1}{2}\sqrt{\frac{k}{15v_{2}}}\approx-\frac{1}{4}\sqrt{\frac{k}{4v_{2}}},
\end{equation}
which is $\epsilon_1\sim 1\,\textit{TeV}^{-1}$ for $k^{3}\sim
v_{2}\sim 1\,\textit{TeV}\,^{3}$. The radion mass in this case can
be of the order of hundreds of $\textit{GeV}$.

\section{Conclusion}
In the present paper we discussed a stabilized brane world model
in five-dimensional Brans-Dicke theory and found equations of
motion for the fields describing excitations above the background
solution. A convenient gauge and a substitution were found
allowing one to diagonalize the equations of motion and to isolate
the physical degrees of freedom of the model. Analogously to the
case of the stabilized model with the minimal coupling of scalar
field to gravity, the tensor sector decouples from the scalar one.
For the background solution (\ref{fon}) we found the mass spectra
of tensor and scalar modes and the coupling constants to matter on
the brane at $y=L$, where our world is assumed to be. It was shown
that the effective four-dimensional Lagrangian does not contain
tachyons and phantom fields. It turned out that, contrary to the
stabilized Randall-Sundrum model, the linearized equations of
motion for the case of background solution (\ref{fon}) can be
solved analytically for all physically interesting values of the
parameters of the model, i.e. this model is exactly solvable in
the linear approximation. This fact can be useful for obtaining
estimates of the influence of the extra dimension on processes on
the brane for different values of parameters of the
five-dimensional theory.

It was shown that for a certain choice of the model parameters the
radion mass may be of the order of hundreds of $GeV$, the inverse
size of the extra dimension and the masses of tensor excitations
being of the order of $TeV$. The coupling constants of massive
tensor and scalar modes appear to be of the order of $TeV^{-1}$,
whereas the coupling constant of the massless graviton appears to
be $\sim M_{Pl}^{-1}$, i.e. the hierarchy problem of gravitational
interaction is solved on the brane at $L$.

Finally we would like to note that the explicit form of the
background solution for functions $\sigma(y)$ è $\phi(y)$ was used
only for calculating the mass spectrum and coupling constants. All
the results related to the gauge choice, the diagonalization of
linearized equations of motion and the structure of tensor and
scalar sectors are valid for any scalar field potential in a
stabilized brane world model in five-dimensional Brans-Dicke
theory with background solution of the form (\ref{backgmetric}),
(\ref{backgscalar}).

\section*{Acknowledgements}
The work was supported by grant of Russian Ministry of Education
and Science NS-1456.2008.2. M.S. acknowledges support of grant for
young scientists MK-5602.2008.2 of the President of Russian
Federation and grant of the "Dynasty" Foundation.


\begin{thebibliography}{99}
\bibitem{Sun}
L. Randall, R. Sundrum, Phys. Rev. Lett 83 (1999) 3370.

\bibitem{Rubakov}
V.A. Rubakov, Phys. Usp. 44 (2001) 871.

\bibitem{Kubyshin}
Yu.A. Kubyshin, arXiv:hep-ph/0111027.

\bibitem{boos}
E.E.~Boos, I.P.~Volobuev, Yu.A.~Kubyshin, M.N.~Smolyakov, Theor.
Math. Phys. 131 (2002) 629.

\bibitem{wolfe}
O. DeWolfe, D.Z. Freedman, S.S. Gubser, A. Karch, Phys. Rev. D62
(2000) 046008.

\bibitem{BVMS}
E.E.~Boos, I.P.~Volobuev, Yu.S.~Mikhailov, M.N.~Smolyakov, Theor.
Math. Phys. 149 (2006) 1591.

\bibitem{Grzadkowski}
B. Grzadkowski, J. .F. Gunion, Phys. Rev. D68 (2003) 055002.

\bibitem{mmsv}
A.S. Mikhailov, Yu.S. Mikhailov, M.N. Smolyakov, I.P. Volobuev,
Class. Quant. Grav 24 (2007) 231.

\bibitem{Aref}
I.Ya. Aref'eva, M.G. Ivanov, W. Muck, K.S. Viswanathan, I.V.
Volovich, Nucl. Phys. B590 (2000) 273.

\bibitem{11}
Ch. Charmousis, R. Gregory, V. Rubakov, Phys. Rev. D62 (2000)
067505.

\bibitem{BKM}
V.M.~Babich, M.B.~Kapilevich, S.G.~Mikhlin, "Linear Equations of
Mathematical Physics", Nauka, Moscow, 1964 (in Russian).
\end{thebibliography}
\end{document}